\documentclass[a4paper,final]{appolb}

\usepackage{amsmath}     
\usepackage{graphicx}    
\usepackage{color}       
\usepackage{subfigure}   
\usepackage{cite}

\begin{document}

\title{\bf 
           Competition of Commodities for the Status of Money in an Agent--based Model
      }
\author{Robert G\c{e}barowski${}^{1}$,
        Stanis\l{}aw Dro\.zd{\.z}${}^{1,2}$,
       Andrzej Z. G\'orski${}^2$ \and 
      Pawe\l{} O\'swi\c{e}cimka${}^2$
\address{${}^1$ Wydzia\l{} Fizyki, Matematyki i Informatyki, ul. Warszawska 24, \\
            Politechnika Krakowska im. Tadeusza Ko\'sciuszki, 31-155 Krak\'ow, Poland; \\
        ${}^2$ Instytut Fizyki J\c{a}drowej PAN, ul. Radzikowskiego 152, 31-342 Krak\'ow, Poland.}       
       }       

\date{\today}

\maketitle

\begin{abstract}
In this model study of the commodity market, we present some evidence of competition of commodities for the status of money 
in the regime of parameters, where emergence of money is possible. 
The competition reveals itself as a rivalry of a few (typically two)
dominant commodities, which take the status of money in turn.
 
\end{abstract}


\section{Introduction}

The world foreign exchange market complex behavior with its money currencies values related to each other poses 
a big challenge for traders, economists and physicists \cite{Lux1999, Drozdz2010, Oswiecimka2006, Yakovenko2009, Kwapien2012}.  
The money origin \cite{Menger1892} phenomenon itself can be successfully reproduced within
 an agent--based computational economics model,
as it has been demonstrated by Yasutomi \cite{Yasutomi1995, Yasutomi2003}. 
In such a model, a number of agents producing different 
types of commodities, exchange them searching for wanted goods which are then consumed. 
The exchange rules are not only governed by individual agent's demands but also rely 
on the market view on each particular commodity. This view is in a way derived from previous exchange transactions
of the ensemble of agents. Thus some commodities may become relatively more desirable, 
becoming widely recognized as an universal mean of exchange for substantial length of time,
 before another commodity overtakes the status of commodity based money in this model market.  

A detailed study of such variant of the agent--based computational model by G\'orski \etal (2010) \cite{Gorski2010}
 discussed further the notion of money and the criteria of money emergence,
 as well as the money switching phenomena, whereby different commodities overtake the dominant role
 on the model market and gain the status of money. Recent study by Dro\.zd\.z \etal (2013) \cite{Drozdz2013} have shown some 
interesting features of the model, which are typical to real financial markets. In particular,
 near the critical threshold, when the onset of the stable phase money in the barter exchange occurs,
 it turns out that time series of fluctuating money lifetimes  exhibits signatures of multiscaling behavior.
Hence the agent--based model, investigated further in the present study, 
is capable to reproduce some of rich and complex behavior of the real financial market.

The goal of present contribution is to study mechanisms of commodity competition for the status of money
 within the considered model. The main research interest will be focused on statistical 
signatures accompanying spontaneous emergence of money and its alternation where just a few commodities,
typically a pair of them, are in rivalry. 

In the following section, we briefly review the model used in our present research. 
Then we present results of the numerical simulations. 
Finally we present summary and draw some conclusions.

\section{The Agent--Based Computational Model}

There are many excellent review papers describing various aspects of agent--based computational models for the economy
(e.g. see \cite{Chakra2011a, Chakra2011b, Samanidu2007, Sornette2014}).

The model used in the present research has been introduced, as far as we could trace it back,
by Yasutomi (1995) \cite{Yasutomi1995}. It has been revisited in details by G\'orski \etal (2010) \cite{Gorski2010}
and also recently investigated by Dro\.zd\.z \etal (2013) \cite{Drozdz2013}, where 
all the technical details of the model implementation are given. 
It has been shown by Dro\.zd\.z \etal (2013) \cite{Drozdz2013}, that such model is mature enough to reproduce
some `stylized' facts \cite{Chakra2011a} about real Commodity or FX Markets by means of statistical derived
market `observables'.

In what follows, we will only review the main model features, which are important in the context of the present research. 

In the model market, there is an ensemble of $N$ nearly identical trading agents and $M=N$ different commodities.
There are some microscopic rules for trading strategies which determine the dynamics and 
statistical ensembles for outcomes (`observables'). 

A generic single transaction consists of a few steps, including: a random choice of an agent, matching a co--trader for
the chosen agent, their interaction by means of exchange one--to--one of their goods according to their preferences
and averaged market opinion, the consumption and the production. A round of $N$ such transactions is called a turn 
and sets a unit of time $t$, allowing for a dynamical change of various microscopic and macroscopic parameters of the model system. 

Therefore the system dynamics will be investigated in fictitious 
time $t$, called the transaction time, measured in 
discrete steps, called turns. 

Trading agents $k=1, 2, \ldots, N$, are equipped with varying in time $t$ an integer number of $j$--th commodity 
($j=1, 2, \ldots, M=N$) denoted by $P_t(j,k)$, randomly changing preferences $W_t(j,k)$ as to the most wanted commodity 
(in our model there is only one wanted commodity by an agent, which is not the supplier of that commodity)
 and evolving views (market opinions) $V_t(j,k)$ on the value of any particular commodity, which are in a way averaged 
over the ensemble of trading (interacting) agents.
Thus values $V_t(j,k)$ contain memory on the past transactions. The matched pair of the agent and its co--trader 
increase their own views on any commodity in case their demand was not satisfied in the previous transaction.
Subsequently they average their mutual views on each particular commodity. Varying in time values $V_t(j,k)$ are normalized with 
respect to the total number of commodity types ($M=N$) in such a way that $ 1 \le  V_t(j,k) \le M$.

The commodity $j$ enters a wish--list of the agent $k$ also when $V_t(j,k) \ge T$. 
That is, once a view on a particular commodity becomes equal or greater than a macroscopic model parameter $T$,  $V_t(j,k) \ge T$,
the commodity $j$ becomes also wanted by a trading agent $k$. 
This represents an external market view on the degree of attractiveness 
of such commodity. Recall that each trader is also driven by the internal (individual and independent) need for a single 
randomly chosen commodity given by $W_t(j,k)$. 

The global macroscopic activation of interest parameter is called the threshold parameter $T$.

At the end of each transaction, the desired commodities are fully consumed (expended) by a trading pair of agents.
If there is no self--supplied (produced) commodity in the portfolio of any of these two agents 
(the agent $k$ delivers the commodity $j$; it
is assumed for simplicity that $j=k$), than a unit of such commodity is produced.

The statistical ensemble for the system is created out of a certain number of random initial conditions for initially
preferred commodity, that is a number of `trajectories' for the system is obtained through dynamics arising
from different $W_0(j,k)$ values in trading time measured in turns.
It is also worth to point out, that for any given randomly chosen initial condition, 
during the system time evolution there is also a stochastic component
due to random choice of trading agents as described above.              

\section{The results}

\subsection{Emergence of the money}

The model allows for a study of the global market view on the strength of a single commodity by means of 
its universality in trading or perceived attractiveness on the commodity exchange market.

\subsubsection{The notion of money}

Let us define the commodity strength $V_{CS}^{(j)}(t)$  
as normalized view on commodity $j$, averaged over the agent's ensemble: 
\begin{equation}
V_{CS}^{(j)}(t) := \left \{ \frac{1}{N}\sum_{k} V_t(j,k) \right \},
\label{vcs}
\end{equation}

Such commodity strength is maximized for a certain $j = j_{max}$:
\begin{equation}
V_{max}(t) := \smash{\displaystyle\max_{j}} \left \{ V_{CS}^{(j)}(t) \right \} 
\label{vmax}
\end{equation}

We say that, the status of money is hence reached by the commodity $j_{max}$.
Therefore the money strength $V_{max}(t)$ is a strength measure of the status of money reached by the commodity $j_{max}$. 

Commodity competition for the status of money would be rivalry of a few commodities to maximize $V_{max} \in [1, N]$.

However one has to remember that reaching status of money in the sense of Eq.~(\ref{vmax}) might be an oversimplified
way to explain the money phenomenon. As discussed earlier (e.g. G\'orski {\it et al}, 2010),
 a detailed study of money emergence requires fulfillment of some additional conditions
(e.g. a relatively long lifetime of the commodity with the status of money). 

\subsubsection{Money strength}

\begin{figure}[htb]
    \centerline{
    \includegraphics[scale=0.45]{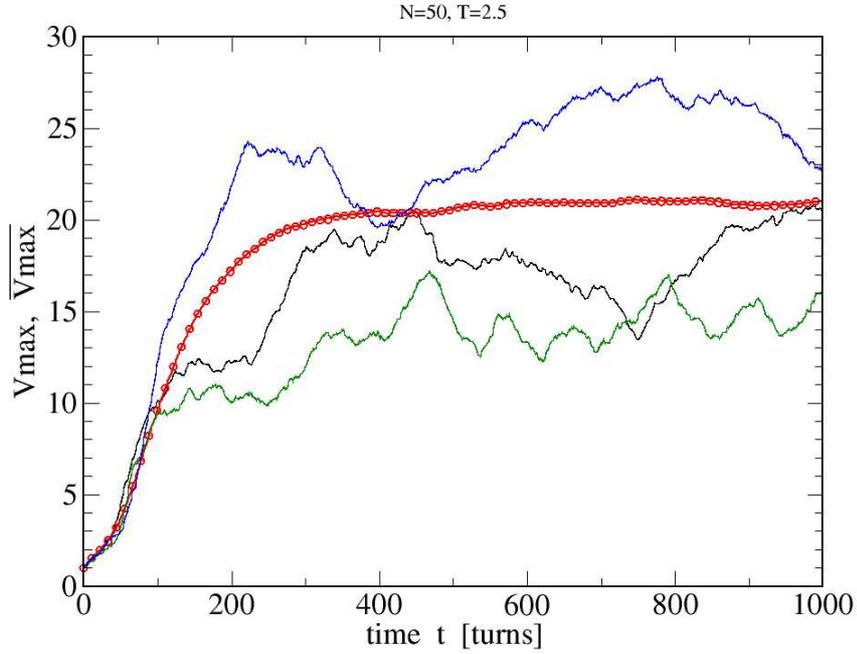}
               }
    \caption{The money strength $V_{max}$ versus time $t$ averaged over $nr=100$ realizations 
              for $T=2.5$ and $N=50$ (a red curve with open dots). There are also shown 3 examples of statistical realizations
             (black, blue and green lines).}
    \label{vmax_time}
  \end{figure}

In order to study the money strength (cf. Eq.~(\ref{vmax})) originating in the model, we made a statistical averaging
over initial conditions (`trajectories'). Figure Fig.~(\ref{vmax_time}) illustrates typical 3 examples of such system `trajectories' 
in time for the threshold parameter $T=2.5$ and $N=50$ agents.  
The smooth--out average $\bar{V_{max}}$, obtained for $nr=100$ realizations, is shown
by the red line with dots. This provides some indication as to the spread of the results for various realizations
 and justifies the need for statistical averaging in order to get some reliable estimates of the money strength. 

\begin{figure}[htb]
    \centerline{
    \includegraphics[scale=0.45]{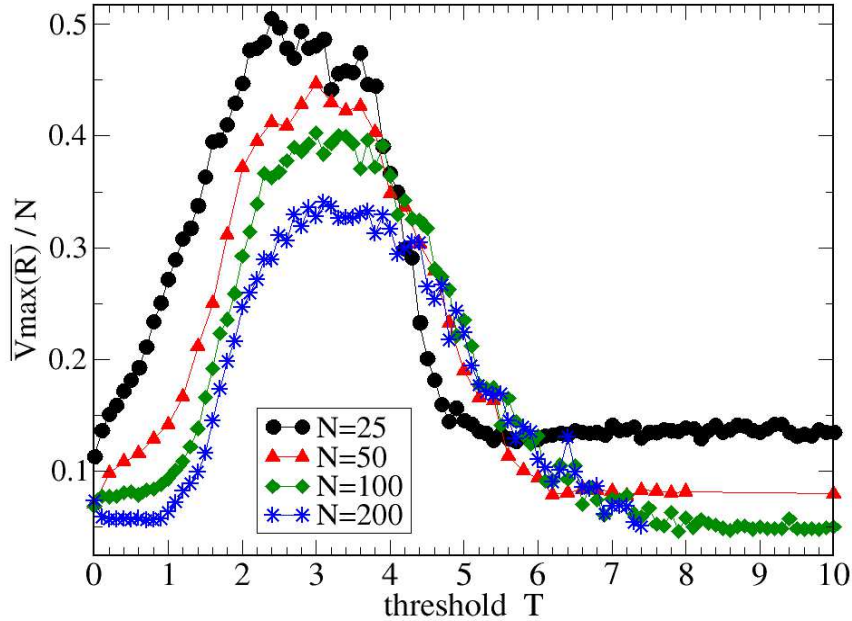}
               }
    \caption{The money strength $V_{max}(R)$ after $t=R=1000$ turns, 
            averaged over $nr=100$ realizations, versus the threshold $T$. 
             Data shown for various numbers of agents in the model: dots -- $N=25$, triangles -- $N=50$, diamonds -- $N=100$
             and asterisks -- $N=200$. 
             Note that, for the purpose of this comparison, the money strengths have been rescaled by the parameter $N$.
   }
    \label{vmax_thresh}
  \end{figure}

In figure Fig.~(\ref{vmax_thresh}) the scaled strength of the money status versus threshold $T$
 is shown for various numbers $N$ of agents taken for the simulations of our model market.
The money strength is averaged for $nr=100$ realizations after $t=R=1000$ turns of the trading time in each case.
The figure shows a general outlook on the money emergence from the  `barter trade' phase to the phase of a `single universal money'
in the sense given by Eq.~(\ref{vmax}). Note also the presence of the `starvation' (the money `collapse') 
phase shown by a small asymptotic value of the money strength \cite{Gorski2010}. 
Noteworthy is that the strong money phase (the shape and position of the maximum)
weakly depends on $N$. This weak dependence is due to a system finite--size effect for a given time of the commodity market simulations.
Nevertheless, the existence of that maximum is quite robust for a range of parameters shown in this model computations. 
Hence, this gives a strong motivation for more insight into the mechanism of gaining the status of money in this regime. 

Therefore one may conclude that, the money strength as defined by Eq.~(\ref{vmax}) 
is a good global indication of the money--phase presence. This is a region in threshold values for 
approximately $ 2 \le T \le 5$,
where one commodity is clearly dominant.

\subsubsection{Money competition}

\begin{figure}[htb]
   \centerline{
   \includegraphics[scale=0.45]{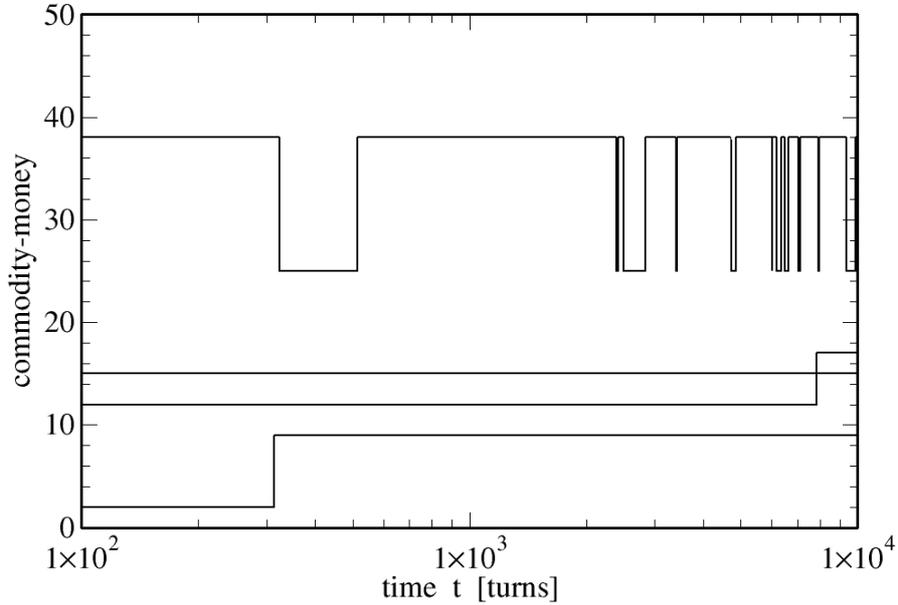}
              }
    \caption{Commodity with the status of money (commodity--money) versus trading time $t$ shown for a time interval 
    ranging over 2 orders of magnitude.
      An example of 4 `trajectories' obtained for $N=50$ and $T=2.5$. Note the case 
       of the commodity competition between two of them 
      in the time interval shown.  A pair of rivalling commodities interchange in the position of the dominant 
      commodity with the status of money.}
  \label{money_jumps}    
  \end{figure}

In figure Fig.(\ref{money_jumps}) the competition for the status of money is exemplified for $N=50$ and $T=2.5$.
A clear interchange of the `money' role is seen among dominant commodities over the range of 2 orders of magnitude
in the time interval shown. This phenomenon
has been referred to as the `money switching' by G\'orski \etal \cite{Gorski2010}. As we can 
see, a given commodity can gain the status of money for a relatively long periods of time
with short time--spans, when a companion (a `runner--up`) commodity takes over. 

It turns out that, for $T \sim 3$ such competition occurs mostly between two commodities only. 
Recall that this value of the threshold is close to the position of the maximum strength of the money--phase
(see Fig.~(\ref{vmax_thresh}). 
It is interesting to note however that, in general such rivalling commodities keep
the money status over different time scales.

\subsubsection{Money lifetimes}

Such phenomenon of money competition occurring in the strong money phase raises 
a question about typical lifetimes of competing commodities.
As we have observed in our simulations, such lifetimes may span over many orders of magnitude. 
The lifetimes could also be very short (cf. Fig.~\ref{money_jumps}) and last for just about a few turns.

Let us consider a given realization (a `trajectory' for fixed $N$ and $T$ values). 
Such a realization provides us with a series of $m$ times, 
$t_1, t_2, t_3, \ldots, t_{m}$,
when the commodity--money changes over the observation time $t_{obs}$. 
This yields a series of lifetimes for any commodity, currently with the status of money over that `trajectory':
$\tau_1 = t_2 - t_1,  \tau_2 = t_3 - t_2,  \ldots, \tau_m = t_m - t_{m-1}$.
            
If we take an ensemble of all `trajectories' (realizations), we obtain
a combined statistics of money lifetimes, that is the lifetimes of the commodity which has 
the status of money.
In order to study commodities in a stable (long--lived) money phase,
we exclude from the distribution lifetimes shorter than $10$ turns
(short--lived money states, lasting just a few turns). 

\begin{figure}[htb]
   \centerline{
   \includegraphics[scale=0.45]{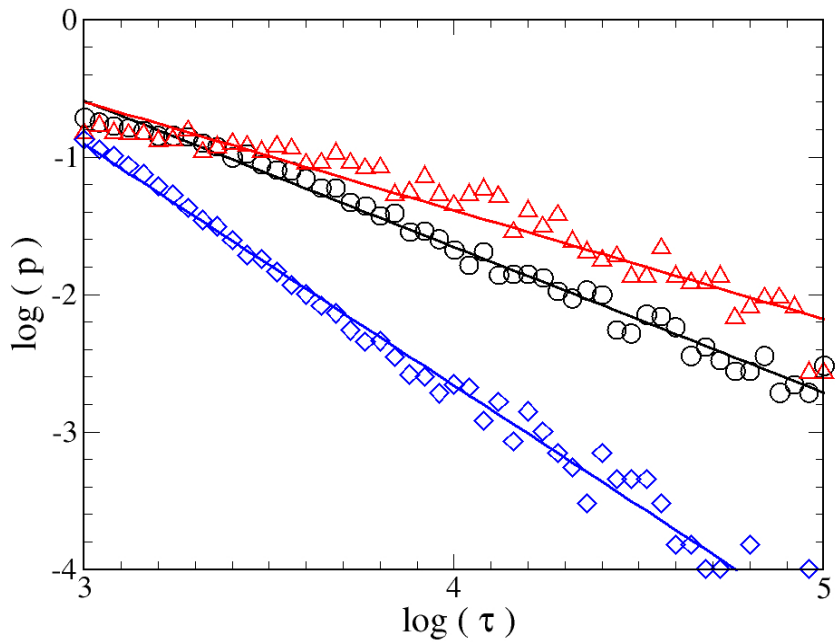}
              }
     \caption{Money lifetimes probability distribution. This is the combined lifetimes' statistics of commodities 
         with the status of money.
          Parameters for the results shown are: $N=50$, time $t_{obs}=1\ 000 \ 000$ turns,  $nr=100$ realizations. 
        Typically $10^4 - 10^5$ lifetimes per histogram were collected.
         The window cut-off for the lifetimes, $1 \le \textrm{log}(\tau) \le 5$. Diamonds are for threshold $T=2.0$, 
      dots for $T=2.5$, and triangles for $T=3.0$. Note the double--log scale on the graph. 
         }
  \label{plog_lifetime}       
  \end{figure}

In figure Fig.(\ref{plog_lifetime}) money lifetime statistics are shown for three different threshold
$T$ parameters corresponding to the vicinity of the maximum in the money strength.
 The numerical results represented with diamonds were obtained for the threshold value $T=2.0$, 
      dots show the data for $T=2.5$, whereas triangles correspond to $T=3.0$. 
Note the double--log scale on this graph. 

The simulations for each `trajectory' were carried out over the observation time of $t_{obs}=1\ 000 \ 000$ turns.
There were $nr=100$ such `trajectories' (realizations) taken for each value of the threshold shown.
Typically $10^4 - 10^5$ lifetimes were collected per histogram. Since we have adopted
 a fixed observation time in the simulations to collect the data, 
we have also rejected from the ensemble lifetimes longer than $10^5$ turns in order to avoid artefacts 
(a premature end to the lifetime of any commodity due to the finite observation time
 would increase the probability of lifetimes near the cut-off value of $10^5$ turns).

Figure (\ref{plog_lifetime}) clearly provides a strong evidence for the power--law behaviour of the tail of that distribution
($\textrm{log}(\tau) > 3$) over the range of at least two orders of magnitude.
 Fitted exponents to the power--law

\begin{equation}
\label{powerlaw}
  p(\tau) \sim A \ \tau ^{-\alpha}, 
\end{equation}

\noindent are the following:
  $\alpha =  1.75$ (for the threshold $T=2.0$), $\alpha = 1.06$ (for $T=2.5$),
and $\alpha = 0.79$ (for $T=3.0$). 
The standard deviations for these values are comparable and approximately equal to $0.03$.
Obtained numerically fits are shown with solid lines in the figure.
This is an interesting additional signature of the money behaviour in the strong money phase, which has been shown
elsewhere using a different approach of
the multifractal singularity spectra analysis \cite{Drozdz2013}.
Moreover, our present results show that
 by choosing an appropriate threshold value, it is possible to investigate a cross--over
of exponents near the value of $\alpha_c =  1.0$ within studied agent--based model. 
By means of a single adjustable parameter, the threshold $T$,
one is able to reproduce different long--tail behaviours of lifetimes of certain processes
 in the model market. This offers an exciting possibility for future
research to investigate real market data with such calibrated agent--based model.

\section{Summary and Conclusions}
 
In this paper we have demonstrated strong competition for the status of money in an agent--based
commodity market model. The competing commodities
have their lifetimes distributed according to the power--law behavior for lifetimes 
longer than $10^3$ turns of trading time. 
Thus the agent--based model is a good testing ground for investigating
rare events statistics. Future research may
also focus on the sensitivity of power--law exponents on details of microscopic strategy adopted
by trading agents. As argued in this paper, the model may also be useful in reproducing some
exponent values of power--law behaviours, observed for real markets.
This could have some possible applications in improving
automatic trading algorithms.

\section*{Acknowledgments}
We would like to acknowledge enlightening discussions with Jaros\l{}aw Kwapie\'n.
One of us (RG) would like to acknowledge the use of the services and computer resources 
provided by the Academic Computer Center
CYFRONET AGH in Krak\'ow  (Akademickie Centrum Komputerowe 
CYFRONET AGH, Grant No. MNiSW/IBM\_{}BC\_{}HS21/PK/033/2014, \hfill\break
``Te\-o\-ria z\l{}o\.zono\'sci rynk\'ow finansowych'').

\end{document}